\documentclass[prstab,amsmath,preprint,showpacs]{revtex4-1}
\usepackage{graphicx,hyperref,siunitx}

\usepackage{lineno}
\usepackage{dsfont}

\newcommand\eps{\varepsilon}

\newcommand\pt{\partial}
\newcommand\beq{\begin{equation}}
\newcommand\eeq{\end{equation}}

\newcommand{\degree}{\mbox{$^{\circ}$}}
\newcommand{\tilvec}[1]{\vec{\tilde{#1}}}

\begin{document}


\title{Equilibrium parameters in coupled storage ring lattices and practical applications}

\author{V. Ziemann}
\email{volker.ziemann@physics.uu.se}
\affiliation{Department of Physics and Astronomy, Uppsala University, Uppsala, Sweden}
\thanks{Orcid ID \url{https://orcid.org/0000-0002-6229-5620} }
\author{A. Streun}
\email{andreas.streun@psi.ch}
\affiliation{Paul Scherrer Institut, CH-5232 Villigen, Switzerland}
\thanks{Orcid ID \url{https://orcid.org/0000-0002-0209-7358} }

\date{\today}

\begin{abstract}
We calculate equilibrium emittances and damping times due to the emission
of synchrotron radiation for coupled storage ring lattices by
evaluating the projections of the commonly used synchrotron radiation
integrals onto the normal modes of the coupled motion.
Orbit distortion is included by calculating off-axis contributions to the radiation integrals.
We provide explicit formulae for fast forward calculation, which have been implemented
into the interactive lattice design code OPA.
\end{abstract}

\pacs{29.20.Dh, 29.27.Bd, 41.60.Ap}

\maketitle

\section{Introduction}
The emittance is one of the crucial quantities that determines the transverse
beam dimensions and thus the performance of storage rings. In electron rings it
is determined by the balance of damping and excitation due to the emission of
synchrotron radiation~\cite{SANDS}. For rings with negligible coupling between the
transverse planes, the effect is quantified by the classical synchrotron radiation
integrals~\cite{HELM,SYPHERS}. Coupling was initially described by a heuristic
emittance coupling constant $\kappa=\eps_y/\eps_x$ where $\eps_x$ and $\eps_y$
are the horizontal and vertical emittance, respectively. With the increased
performance of accelerators elaborate methods to calculate the emittance
coupling were developed~\cite{SLIM,OHMI,WOLSKI}. We complement these
often elegant, but sometimes less intuitive methods, by a generalization of
the radiation integrals to a coupled lattice.
\par
The beam optics code OPA~\cite{OPA}, written and maintained by one of the authors, is tailored to interactively design beam optical systems. Especially for the initial design phase, estimating the equilibrium beam properties from the optics of a non-periodic system, helps to guide the optimization; and this requires the evaluation of the radiation integrals, especially for intrinsically coupled lattices, such as Spiral-COSAMI~\cite{SPIRAL}.
\par
We consider the normal mode parametrization of the coupled one-turn transfer
matrix as introduced by Edward and Teng ~\cite{EDTENG} and extended by Sagan
and Rubin~\cite{SAGAN}, and calculate the projection of the radiation integrals
onto the corresponding normal modes.
Note that we also implemented the algorithm in MATLAB using the software from
\cite{VZMATLAB} which is useful to validate and cross-check the software.
\par
In the discussion and also with regard to practical implementation, we assume
that all linear beam optical elements such as quadrupoles or dipoles are
defined by their transfer matrices for the un-coupled elements and sandwiched
between coordinate rotation matrices that describe their misalignment pertaining
to coupling of the transverse planes (i.e. a skew quadrupole is represented by
a quadrupole sandwiched between $\pm45\degree$ rotations).
\par
We only consider transverse coupling and assume the longitudinal dynamics to be decoupled
by adiabatic approximation, i.e. the synchrotron oscillation are considered slow compared
to the betatron oscillations, and a particle's longitudinal momentum is assumed as a constant.
Furthermore our analysis is based on highly relativistic, paraxial and large bending radius
approximations as commonly used for high energy storage rings, where beam properties are
determined by synchrotron radiation effects.
\par
We start with a brief reminder of the required beam dynamics formalism, then we calculate the
radiation integral $I_5$ (in the notation of ref~\cite{HELM}) responsible for excitation of
betatron oscillations followed by the evaluation of $I_4$ that describes damping. From these
follow the equilibrium beam parameters of a coupled lattice. We proceed further including  contributions from orbit excursions, which may be due to energy offsets or lattice imperfections. Finally we discuss implementation issues and present some applications.

\section{Normal modes and dispersions}
Transverse beam dynamics using coordinates $(x,x',y,y',\Delta E/E)$ is covered by $5\times 5$
transfer matrices for elements and [circular] concatenations of elements:
\beq\label{eq:rmatrix}
\left( \begin{array}{c|c} R & \vec{V} \\ 0 & 1 \end{array} \right)
\eeq
with $R$ the $4\times 4$ transverse transfer matrix and $\vec{V}$ the 4-dimensional vector of dispersion production.\\
\par
If $R$ is the one-turn matrix of a storage ring, starting at an arbitrary longitudinal position, a normal mode decomposition may be
found~\cite{EDTENG,SAGAN}:
\begin{equation}\label{eq:edteng}
R= T^{-1} {\cal A}^{-1} {\cal O} {\cal A} T
\end{equation}
where $\cal O$ is the matrix containing the eigen-tunes
\begin{equation}
{\cal O} =\left(\begin{array}{cc} O_a & 0 \\ 0 & O_b\end{array}\right)
\quad\mathrm{with}\quad
O_{a,b}=\left(\begin{array}{rc} \cos\mu_{a,b} & \sin\mu_{a,b} \\ -\sin\mu_{a,b} & \cos\mu_{a,b}\end{array}\right)
\end{equation}
and ${\cal A}$ is the matrix that contains the local normal-mode beta functions. It is given by
\begin{equation}\label{eq:calA}
{\cal A} = \left(\begin{array}{cc} {\cal A}_a & 0 \\ 0 & {\cal A}_b\end{array}\right)
\quad\mathrm{with}\quad
{\cal A}_{a,b} =
\left(\begin{array}{cc} \frac{1}{\sqrt{\beta_{a,b}}} & 0 \\ \frac{\alpha_{a,b}}{\sqrt{\beta_{a,b}}} & \sqrt{\beta_{a,b}}
\end{array}\right)
\end{equation}
where the indices $a$ and $b$ label the two eigen modes.
The coupling matrix $T$ and its inverse $T^{-1}$ are given by~\cite{SAGAN}
\beq\label{eq:tmatrix}
T=\left( \begin{array}{cc} gI & -C \\ C^+ & gI \end{array} \right)
\qquad \mathrm{and}\qquad
T^{-1}=\left( \begin{array}{cc} gI & C \\ -C^+ & gI \end{array} \right) ,
\eeq
with the $2\times 2$ identity matrix $I$ and the $2\times 2$ coupling matrix $C$.
Its symplectic conjugate $C^{+}$ and the scalar $g$ are given by
\beq
C^+=C^{-1}\det C ,\quad \mbox{\rm and}\quad  g^2=1-\det C.
\eeq

\par
The periodic solution for the four-dimensional dispersion
$\vec D=(\vec D_x, \vec D_y )^T= (D_x,D'_x,D_y,D'_y)^T$ is constrained by
\begin{equation}
\vec D = R \vec D + \vec V
\end{equation}
which can be solved for $\vec D$ with the result
\begin{equation}\label{eq:disp}
\vec D = (1-R)^{-1}\vec V\ .
\end{equation}
Thus $\vec D$ describes the periodic dispersion in physical space.
The normal mode dispersion is given by
\begin{equation}\label{eq:nmodedisp}
\vec{\cal D} = (\vec{\cal D}_a, \vec{\cal D}_b)^T=
({\cal D}_a,{\cal D}_a',{\cal D}_b,{\cal D}_b')^T= T\vec{D}.
\end{equation}
Further application of matrix $\cal{A}$ gives the normal mode dispersion in normalized phase space:
\begin{equation}\label{eq:Dtilde}
\tilvec{D}=
(\tilde{D}_1, \tilde{D}_2, \tilde{D}_3, \tilde{D}_4)^T=
{\cal A}\vec{\cal D} = S^{-1}\vec D
\end{equation}
\beq\label{eq:smatrix}
\mbox{\rm with~~~~} S=({\cal A} T)^{-1} = T^{-1}{\cal A}^{-1}.
\eeq
We thus defined dispersion in three different coordinate systems, which we will use as it is convenient, and which are related as
\[
\tilvec{D}_{\mbox{\rm ~(normalized)}}
\quad\stackrel{\cal A}{\longleftarrow} \quad
\vec{\cal D}_{\mbox{\rm ~(decoupled)}}
\quad\stackrel{T}{\longleftarrow}\quad
\vec{D}_{\mbox{\rm ~(real space)}}
\]

Having laid out the theoretical framework and definition of notations we proceed to calculate the effect of quantum excitations on the normal modes.
\section{Quantum excitation}
\label{sec:ex}
The stochastic nature of the emission of synchrotron radiation causes heating
of the beam if the photons are emitted at locations with dispersion. In a planar
un-coupled lattice the effect of the lattice parameters, such as beta function
$\beta$ and dispersion $D$ is described by
\begin{equation}\label{eq:Hflat}
{\cal H} = \beta_x D_x^{\prime 2} + 2\alpha_x D_x D_x' + \gamma D_x^2
\end{equation}
with $\gamma=(1+\alpha^2)/\beta.$ The synchrotron radiation integral $I_5$ is
closely related to ${\cal H}$ by
\beq
I_5=\oint {\cal H} \vert h\vert^3\,ds\ ,
\eeq
where $h = 1/\rho$ is the local curvature in the bending magnets, and the integral extends
over one turn in the ring as discussed in chapter~7 of~\cite{SYPHERS}.
We follow the same logic to generalize the derivation to the
coupled case and assume to know the dispersion in real space coordinates, $\vec D$ at every point in the ring and especially inside
the dipoles, where we know that the emission of photons happens. This emission is
characterized by an average loss, which causes damping and a fluctuating part $u$
that has average energy value zero $\langle u\rangle=0$ and squared expectation
value~$u^2.$ Here $u=\Delta E/E$ is the relative energy deviation of a particle.
We consider only one location and in a particular realization of the random process
the energy loss at a given turn $i$ will be denoted by $u_i.$ We will use the
statistical properties later on. In order to understand the excitation process we
consider a single electron that travels on the closed orbit appropriate for its
energy. After the emission process, the electron is still at the same place, but the
orbit appropriate for its new energy is different, because of the finite dispersion
at the location of emission. Consequently the electron will start oscillation around
the new equilibrium orbit. The change of coordinates $\Delta\vec{x}$ at which it starts the oscillation are given by
\begin{equation}
\Delta\vec x = -u_i \vec D
\end{equation}
and after traveling $n$ turns with the chance of an emission process on each turn we
expect the electron's position to be
\begin{equation}\label{eq:xn}
\vec x_n = -\sum_{i=1}^n R^{n-i} u_i \vec D\ .
\end{equation}
Using the parametrization from eq.~\ref{eq:edteng} it is obvious that powers of the transfer
matrix $R^m$ can be evaluated by
\begin{equation}
R^m = T^{-1} {\cal A}^{-1} {\cal O}^m {\cal A} T
\end{equation}
which can be interpreted in the following way. Reading from right to left: $T$ removes the coupling,
${\cal A}$ turns the ellipses into circles, and then ${\cal O}$ takes care of the phase advance or
tune for the specified number of turns $m.$ Finally we move back from normalized phase space to
real space by the inverses of ${\cal A}$ and~$T.$
\par
Returning to eq.~\ref{eq:xn} we simplify it by mapping the entire equation into normalized
phase space by left multiplying with ${\cal A}T$. We use the notation $\tilvec{x}={\cal A}T\vec x$
and write
\begin{equation}\label{eq:multiturn}
\tilvec{x}_n = -{\cal A}T \left( \sum_{i=1}^n R^{n-i} u_i\right) \vec D
= -\left( \sum_{i=1}^n {\cal O}^{n-i} u_i\right){\cal A}T\vec D
=-\left( \sum_{i=1}^n {\cal O}^{n-i} u_i\right)\tilvec{D}
\end{equation}
with $\tilvec{D}$ the dispersion vector mapped into normalized phase space from eq.~\ref{eq:Dtilde}.
Equation~\ref{eq:multiturn} has a nice intuitive interpretation. Instead of every
energy loss $u_i$ producing a kick of magnitude $u_i \vec D$ in real space, it produces
a kick of magnitude $u_i \tilvec{D}$ in normalized phase space that propagates by multiplying it
with the rotation matrix ${\cal O}$ for a given number of turns and eventually all the
kicks from the separate turns are summed up.
\par
In order to calculate the emittance growth due to such a sequence of kicks we first need
to specify which emittance we really mean. In the following we will use the emittances
of the two normal modes. In the case of an uncoupled lattice this will revert to the common
definition of emittances. Moreover, the emittances are the ensemble averages over the Courant-Snyder
action variables $J_a$ and $J_b$ for the respective normal modes, labeled $a$
and~$b.$ We start by writing the previous equation in $2\times 2$ block matrix form
\begin{equation}\label{eq:txn}
\left(\begin{array}{c} \tilvec{x}_{a,n} \\ \tilvec{x}_{b,n} \end{array}\right)
= -\sum_{i=1}^n u_i
\left(\begin{array}{cc}O_a^{n-i}& 0 \\ 0 & O_b^{n-i}\end{array}\right)
\left(\begin{array}{c} \tilvec{D}_a \\ \tilvec{D}_b \end{array}\right)
= -\sum_{i=1}^n u_i
\left(\begin{array}{c} O_a^{n-i}\tilvec{D}_a \\ O_b^{n-i}\tilvec{D}_b \end{array}\right)
\end{equation}
where $\tilvec{x}= (\tilvec{x}_a, \tilvec{x}_b)^T.$ Here $\tilvec{x}_a$ are just the first two
components of the vector $\tilvec{x}$ and $\tilvec{x}_b$ components three and four.
The action variable for the first normal-mode after $n$ turns is thus given by
\begin{equation}
J_{a,n}=\frac{1}{2}\tilvec{x}_{a,n}^T \tilvec{x}_{a,n} \ .
\end{equation}
We now insert $\tilvec{x}_{a,n}$ from eq.~\ref{eq:txn} and arrive at
\begin{eqnarray}
J_{a,n}
&=& \frac{1}{2} \sum_{i=1}^n\sum_{j=1}^n u_i u_j
\tilvec{D}_a^T \left(O_a^{n-j}\right)^T O_a^{n-i} \tilvec{D}_a\nonumber\\
&=& \frac{1}{2} \sum_{i=1}^n\sum_{j=1}^n u_i u_j
\tilvec{D}_a^T \left(O_a^T\right)^{n-j} O_a^{n-i} \tilvec{D}_a\\
&=& \frac{1}{2} \tilvec{D}_a^T \sum_{i=1}^n\sum_{j=1}^n O_a^{j-i} u_i u_j\tilvec{D}_a
\nonumber
\end{eqnarray}
where we use that the power of a transpose matrix equals the transpose of the matrix
to the same power. Furthermore, the matrix $O_a$ is a rotation matrix and therefore
orthogonal. This implies that its transpose equals its inverse.
\par
Now we consider the ensemble average over many particles. Since the emission of
photons from turn to turn is uncorrelated and has rms magnitude $u^2$ as discussed before
we have $\langle u_i u_j\rangle =u^2\delta_{ij}$ which allows us to evaluate one sum and the
other one sums over a constant $u^2.$ We obtain
\begin{equation}
\langle J_{a,n} \rangle = \frac{1}{2} \tilvec{D}_a^T \sum_{i=1}^n\sum_{j=1}^n u^2\delta_{ij} O^{j-i}_a\tilvec{D}_a
=  \frac{1}{2} \tilvec{D}_a^T \tilvec{D}_a  n u^2
\end{equation}
and observe that the average action variable grows linearly with the number of turns $n$.
\par
The magnitude of the term $u^2$ is the rms energy kick received by the beam. It is then given
by the ratio of the rms spread of the emitted photons $\dot N \langle \eps^2\rangle$ and the
beam energy $E$ and is given by~(Section 3.1.4 in \cite{HANDBOOK})
\begin{equation}
u^2 = \dot N\langle\eps^2\rangle = \frac{55}{24\sqrt{3}}\frac{\eps_cP_{\gamma}}{E^2}
   = \frac{4c r_e}{3}C_q \frac{\gamma^5}{\vert\rho^3\vert}
\end{equation}
with $r_e$ being the classical electron radius, the constant $C_q$, and the critical photon energy
$\eps_c$ given by
\beq
C_q=\frac{55\, \hbar c}{32 \sqrt{3} \, mc^2}
\qquad\mathrm{and}\qquad
\eps_c=\frac{3\,\hbar c\gamma^3}{2\,\vert\rho\vert}
\eeq
The radiated power is given by
\begin{equation}\label{eq:Pgamma}
P_{\gamma} = \frac{e^2c^3}{2\pi}C_{\gamma} E^2 B^2 = \frac{cC_{\gamma}}{2\pi}\frac{E^4}{\rho^2}\qquad
\mbox{\rm with} \quad C_{\gamma}=\frac{4\pi r_e}{3(mc^2)^3}\ .
\end{equation}
Integrating the action variable $J_a$ over the
circumference of the ring we collect the contributions of all the dipole magnets in
the ring. This allows us to define a growth rate for the normal-mode action variable
\begin{equation}\label{eq:heating}
\frac{d\langle J_a\rangle}{dt}
=\frac{2}{3}C_q r_e \gamma^5 \oint \frac{\tilvec{D}_a^T \tilvec{D}_a}{\vert\rho^3\vert} ds
=\frac{2}{3}C_q r_e \gamma^5 I_{5a}
\end{equation}
and similarly for the other normal mode $b.$ Here we implicitly define the radiation integral
$I_{5a}$ for the first normal mode.
Comparing with the expressions in refs.~\cite{SYPHERS,HANDBOOK,SYLEE} we observe that
${\cal H}_a=\tilvec{D}_a^T \tilvec{D}_a$
takes the place of the commonly used symbol ${\cal H}$ that occurs in the description of
emittance growth in uncoupled rings. It is important to note, that only the absolute curvature
of the bending field is used but not the direction of deflection, because it provides the energy
fluctuations as the source of emittance, and the normal mode dispersions project this noise
on to the transverse eigen-modes.
\par
For an uncoupled lattice the coupling matrix $T$ from eq.~\ref{eq:tmatrix} is unity and the
matrix ${\cal A}$ contains the conventional un-coupled beta functions. Using the first
two components of eq.~\ref{eq:Dtilde} we can write
\begin{eqnarray}
\left(\begin{array}{c} \tilde D_x \\ \tilde D '_x \end{array}\right)
&=&{\cal A}_x \left(\begin{array}{c} D_x \\ D'_x \end{array}\right)
=\left(\begin{array}{cc}
1/\sqrt{\beta_x} & 0 \\ \alpha_x/\sqrt{\beta_x} & \sqrt{\beta_x}
 \end{array}\right)
\left(\begin{array}{c} D_x \\ D'_x \end{array}\right)\nonumber\\
&=&\left(\begin{array}{c}
D_x/\sqrt{\beta_x}\\ \alpha_xD_x/\sqrt{\beta_x} + \sqrt{\beta_x} D'_x
\end{array}\right)
\end{eqnarray}
and multiplying with its transpose, ${\cal H}_x=\tilvec{D}_x^T \tilvec{D}_x$, yields eq.~\ref{eq:Hflat}.
\section{Damping}\label{sec:damp}
The second consequence of the emission of synchrotron radiation is the damping of transverse oscillations.
In dipole magnets synchrotron radiation is emitted and that emission is energy
dependent with the recoil from the photons along the direction of motion of the electrons.
Subsequently the energy is restored in a radio-frequency cavity, but only the longitudinal
component of the momentum vector is increased. The joint effect of emitting and restoring the
energy leads to overall damping. We thus proceed to determine the damping times of the two
normal modes $a$ and $b$ and loosely follow the description from ref.~\cite{SYLEE}.
\par
\begin{figure}[htb]
\centering
\includegraphics[width=70mm]{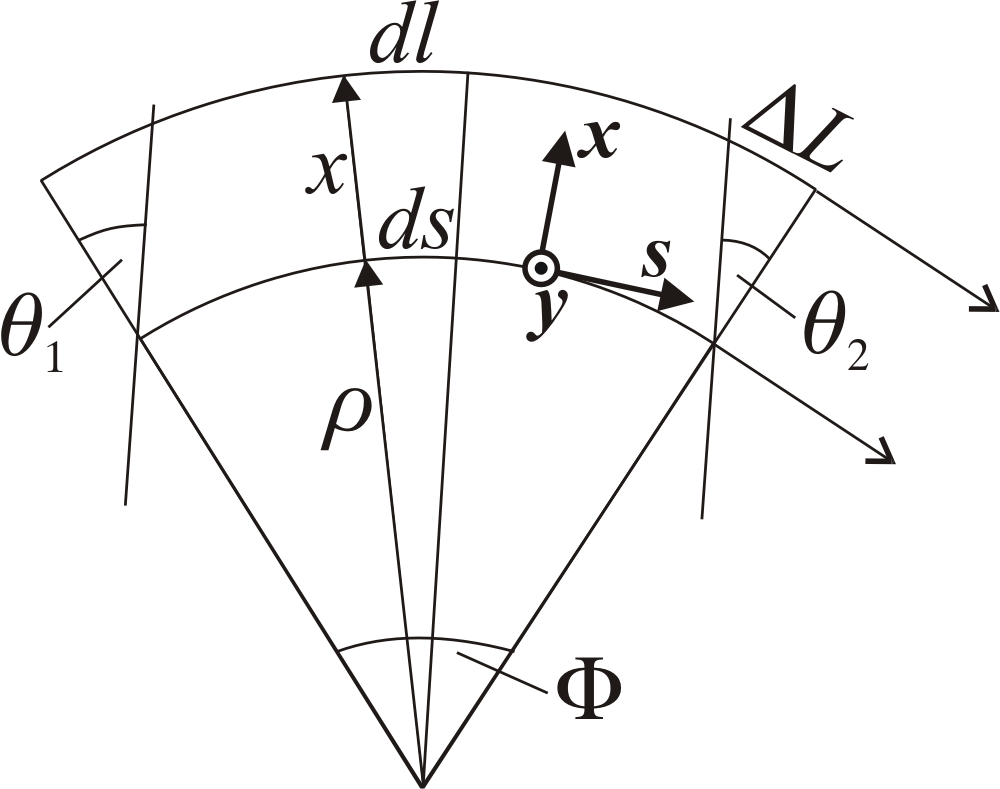}    
\caption{\label{fig:bendgeo}Bending magnet geometry}
\end{figure}
In dipoles we denote the average emitted energy by $\Delta\bar u$ which is related to the power
of the emitted synchrotron radiation $P_{\gamma}$ by $\Delta\bar u=P_{\gamma} dl/cE$ where $dl/c$
is a segment of the arc an electron travels in a magnet, see fig.~\ref{fig:bendgeo}, and $P_{\gamma}$
is given in eq.~\ref{eq:Pgamma}.
\par
If an electron suddenly changes its energy by an amount $\Delta\bar u$ due to photon emission, at a location with dispersion $\vec D$, its reference orbit will jump away from the electron and it will perform betatron oscillations starting with $\Delta\vec x = -\vec D \Delta\bar u$ around the new reference orbit. Multiplying both
$\Delta \vec x$ and the dispersion $\vec D$ by ${\cal A} T$ we obtain the effect on the normal modes
\begin{equation}
\Delta \tilvec{x} = -\tilvec{D} \Delta\bar u
\end{equation}
and the action variables $J_a$ and $J_b$ will change by
\begin{eqnarray}
\Delta J_a &=& \tilde x_1 \Delta \tilde x_1 + \tilde x_2 \Delta \tilde x_2
=  -\left( \tilde x_1\tilde D_1 +\tilde x_2\tilde D_2\right) \Delta\bar u,
\nonumber\\
\Delta J_b &=& \tilde x_3 \Delta \tilde x_3 + \tilde x_4 \Delta \tilde x_4
=  -\left( \tilde x_3\tilde D_3 +\tilde x_4\tilde D_4\right) \Delta\bar u.
\label{eq:jabdefinition}
\end{eqnarray}
We now need to determine the dependence of the energy loss $\Delta\bar u$ on the position in a
combined function magnet with position-dependent magnetic field. We only consider an upright
combined function magnet with only a horizontal dependence of the vertical field component. Other
orientations or rotated magnets can easily be accommodated by coordinate rotations which
enter the analysis by their influence on the coupling matrix $T.$ Thus, without loss of generality,
we describe the total average energy of the emitted photons by~\cite{SYLEE}
\begin{eqnarray}\label{eq:ebarsector0}
\Delta\bar u &=& -\frac{P_{\gamma}}{E}\frac{dl}{c}
= -\frac{P_{0}}{cE}\left(1+\frac{2}{B}\frac{\pt B}{\pt x} x\right)
\left(1+\frac{x}{\rho}\right) ds \\
\label{eq:ebarsector}
 &\approx& -\frac{P_{0}}{cE}\left(1+\frac{2}{B}\frac{\pt B}{\pt x} x + \frac{x}{\rho} \right) ds
\end{eqnarray}
where $x$ denotes the horizontal coordinate in the lab system, and $P_{0}$ is the power
radiated on the design orbit. The term with the derivative of the magnetic field arises from
the quadratic dependence of the emitted power on the magnetic field, see eq.~\ref{eq:Pgamma},
and the second term with $x/\rho$ describes the longer path of the electron in the magnet if
it is further out with larger $x$, see fig.~\ref{fig:bendgeo}.
In dipoles containing also a skew gradient we would need to add $(dBy/dy)y$ to $(dBy/dx)x$ and consistently carry through all following steps.
\par
Expressing the horizontal coordinate $x$ through the normalized normal mode
coordinates $\tilde x$ by using the matrix $S$ from eq.~\ref{eq:smatrix},
\begin{equation}
\label{eq:sum1i}
x = \sum_{i=1}^4 S_{1i}\tilde x_i
\end{equation}
allows us to write the change in the action variables as
\begin{eqnarray}\label{eq:jasector}
\Delta J_a &=&
\frac{P_{0}}{cE}\left(1 +
\sum_{i=1}^4 \left( \tilde x_1\tilde D_1 +\tilde x_2\tilde D_2\right)
 S_{1i}\left( \frac{2}{B}\frac{\pt B}{\pt x}
\tilde x_i + \frac{\tilde x_i}{\rho} \right)\right) ds.
\end{eqnarray}
Averaging over the normal-mode phases denoted by angle brackets, we find
\begin{eqnarray}
\langle\Delta J_a\rangle &=& \frac{P_{0} ds}{cE} \left(\frac{2}{B}\frac{\pt B}{\pt x} + \frac{1}{\rho} \right)
\left(S_{11} \langle\tilde x_1^2\rangle \tilde D_1 + S_{12} \langle\tilde x_2^2\rangle \tilde D_2\right)\, ,
\end{eqnarray}
where we use the fact that $\langle\tilde x_i\rangle = 0$ and $\langle\tilde x_i\tilde x_j\rangle=0$
with $i\ne j$, since the phases are evenly distributed for a large number of particles.
Moreover, the expressions $\langle\tilde x_1^2\rangle = \langle\tilde x_2^2\rangle = \langle J_a\rangle$ are
valid for normalized coordinates such that we finally obtain
\begin{eqnarray}
\langle\Delta J_a\rangle &=& \frac{P_{0} ds}{cE} \left(\frac{2}{B}\frac{\pt B}{\pt x} + \frac{1}{\rho} \right)
\left[S_{11} \tilde D_1 + S_{12} \tilde D_2\right] \langle J_a\rangle \nonumber\\
\langle\Delta J_b\rangle &=& \frac{P_{0} ds}{cE} \left(\frac{2}{B}\frac{\pt B}{\pt x} + \frac{1}{\rho} \right)
\left[S_{13} \tilde D_3 + S_{14} \tilde D_4\right] \langle J_b\rangle \, ,
\end{eqnarray}
where we recover the uncoupled case by setting $S_{11}=1$ and all other $S_{1i}$ to zero. The
expressions with the square brackets and the dispersions $\tilde D$ are the projections of the
coupled dispersions on the normal modes and thus intuitively generalize the uncoupled formalism
to the transversely coupled case. The first two elements of the first row of the matrix $S$ from
eqs.~\ref{eq:calA}--\ref{eq:smatrix} can be evaluated to be $S_{11}= g\sqrt{\beta_a}$ and
$S_{12}= 0$ such that we have
\beq
\label{eq:s12d}
S_{11} \tilde D_1 + S_{12} \tilde D_2 = g{\cal D}_a\ .
\eeq
with ${\cal D}_a$ from eq.~\ref{eq:nmodedisp}.
Since $S$ maps back from normalized phase space to real space eq.~\ref{eq:sum1i} applies to the
physical dispersion as well, and we have
\beq\label{eq:s34d}
D_x =\sum_{i=1}^4 S_{1i}\tilde{D}_i = g {\cal D}_a +S_{13} \tilde D_3 + S_{14} \tilde D_4 .
\eeq
${\cal D}_a$ is the (non-normalized) $a$-mode dispersion, the $b$-mode dispersion is not needed.
We use eq.~\ref{eq:s12d} immediately to simplify the equation for $\langle\Delta J_a\rangle$:
we insert $P_{\gamma}$ from eq.~\ref{eq:Pgamma}
and the definitions of curvature and focusing strength, $h=1/\rho=B/(B\rho)$ and $k=(\pt B/\pt x)/(B\rho)$,
where $k>0$ identifies a horizontally focusing magnet. 
\beq\label{eq:jae4}
\frac{\langle\Delta J_a\rangle}{\langle J_a\rangle} = \frac{C_{\gamma}E^4}{2\pi} \frac{1}{E} g{\cal D}_a (2 kh +h^3)\, ds\, .
\eeq
We proceed to express the constant $C_{\gamma}E^4/2\pi$ by the
total emitted power $U_0$ and the second radiation integral, defined as $I_2=\int h^2 ds$
\begin{equation}
U_0 = \frac{C_{\gamma}E^4}{2\pi} \oint h^2 \,ds = \frac{C_{\gamma}E^4}{2\pi} I_2
\end{equation}
and arrive at
\begin{eqnarray}\label{eq:dampa}
\frac{\langle\Delta J_a\rangle}{\langle J_a\rangle} = \frac{U_0}{E}\frac{1}{I_2} g{\cal D}_a (2 kh +h^3)\, ds
\end{eqnarray}
for the first normal mode. Considering eq.~\ref{eq:s34d}, the relative damping of the second normal
mode amplitude is conveniently expressed by the damping of the first mode amplitude and the
damping in the uncoupled case.
\beq
\label{eq:dampb}
\frac{\langle\Delta J_b\rangle}{\langle J_b\rangle} = \frac{\langle\Delta J_x\rangle}{\langle J_x\rangle} -\frac{\Delta J_a}{J_a} \quad\mbox{\rm with}\quad
\frac{\langle\Delta J_x\rangle}{\langle J_x\rangle} = \frac{U_0}{E}\frac{1}{I_2} D_x (2 kh +h^3)\, ds
\eeq
where $D_x$ is the first component of the real-space dispersion from eq.~\ref{eq:disp}.\\
\par
So far we have considered sector dipoles such that damping of betatron amplitudes is
given by integrating eqs.~\ref{eq:dampa}--\ref{eq:dampb} over the length of the magnet. If, on the other
hand, a magnet edge is rotated by an angle $\theta$ relative to the edge of a pure sector magnet,
with $\theta>0$ approaching a rectangular bend (i.e. a bend where both edge angles are half the bend
angle: $\theta_{1}=\theta_{2}=\Phi/2$), the path of a particle at position $x$ is shortened by
$\Delta L= -x \tan\theta$, see fig.~\ref{fig:bendgeo}, thus its energy loss is reduced compared to the energy loss of a particle on the design orbit by
\beq \label{eq:dledge}
\Delta \bar{u}
= -\frac{P_{0}}{cE} \left( 1+2k\rho x \right) \Delta L
\approx \frac{P_{0}}{cE} x \tan\theta\ .
\eeq
The corresponding change in the action variable is (cf. eq.~\ref{eq:jasector})
\beq
\Delta J_{a} = -\frac{P_{\gamma}}{cE}  ( \tilde{x}_1 \tilde{D}_1
+ \tilde{x}_2\tilde{D}_2) \sum_{i=1}^4 S_{1i} \tilde{x}_i  \tan\theta.
\eeq
Repeating the manipulations from eq.~\ref{eq:jasector} to \ref{eq:dampa} gives
\beq
\label{eq:dampaedge}
\frac{\langle\Delta J_a\rangle}{\langle J_a\rangle} = - \frac{U_0}{E}\frac{1}{I_2} g{\cal D}_a h^2 \tan\theta\ .
\eeq
\par
The relative change of the action variable in a combined function magnet is obtained by
integrating eq.~\ref{eq:dampa} along the design orbit and adding the edge-effects from eq.~\ref{eq:dampaedge}.
After summing over all bending magnets in the lattice we write
\beq\label{eq:damp}
\frac{\langle \Delta J_{a,b}\rangle}{\langle J_{a,b}\rangle} = \frac{U_0}{E}\frac{I_{4{a,b}}}{I_2}\, ,
\eeq
where we have introduced two variants of the fourth radiation integral $I_{4a}$ and $I_{4b}$, one for each normal mode, which are given by an integral over all bending sectors and a sum over all edges:
\begin{eqnarray}\label{eq:intfour}
I_{4a} & = & \oint g(s) {\cal D}_a(s)  (2h(s)k(s)+h(s)^3)\, ds - \sum_k^{\rm edges} g_k {\cal D}_{ak} h_k^2 \tan\theta_k \\
\nonumber
I_{4b} & = & I_4 - I_{4a}\ . \nonumber
\end{eqnarray}
$I_4$ is the well-known integral without coupling, and given by $I_{4a}$ for $g=1$:
\[
I_4 = \oint D_x(s) (2k(s)h(s)+h(s)^3)\,ds - \sum_k^{\rm edges}D_{xk} h_k^2 \tan \theta_k
\]
which reproduces the well-known case~(Section 3.1.4 in \cite{HANDBOOK}). For a small rectangular magnet ($\theta_1=\theta_2=hL/2$) with constant field and length $L$, where the optical functions do not change much over the magnet, eq.~\ref{eq:intfour} simplifies to
$\Delta I_{4a}  \approx  2 g {\cal D}_a  h k L \, .$
\par
The second, and dominant, contribution to transverse damping comes from the restoration of the
longitudinal momentum in the radio-frequency cavities. There, on average, the momentum corresponding to the total energy lost in one turn $\Delta p= -U_0/c$ is added to the longitudinal momentum of the electron, whereas the transverse momenta remain unchanged. Since the transverse angles are the ratio of transverse to the much larger longitudinal momentum, we find~\cite{SYLEE} that both transverse angles are reduced by $-U_0/E.$ This effect affects both transverse coordinates equally and therefore also the coordinates of normalized phase space $\tilde x$, and we have
\begin{equation}
\Delta \tilde x_2 = -\tilde x_2 \frac{U_0}{E}
\qquad\mathrm{and}\qquad
\Delta \tilde x_4 = -\tilde x_4 \frac{U_0}{E}\, ,
\end{equation}
which causes a change in the action variables $\Delta J_a$ and $\Delta J_b$ that, upon averaging
over betatron phases, leads to
\begin{equation}\label{eq:taunull}
\left.\frac{\langle \Delta J_a\rangle}{\langle J_a\rangle}\right\vert_{RF} = \left.\frac{\langle \Delta J_b\rangle}{\langle J_b\rangle}\right\vert_{RF}
= -\frac{U_0}{E}
\end{equation}
for the relative variation of the action variable due to the acceleration in the cavities.
\par
Combining the effect of acceleration and emission of synchrotron radiation from eqs.~\ref{eq:dampa},
\ref{eq:dampb} leads to the following expressions
\begin{eqnarray}
\left.\frac{\langle \Delta J_a\rangle}{\langle J_a\rangle}\right\vert_{tot}
&=& -\frac{U_0}{E} \left(1 - \frac{I_{4a}}{I_2}\right)\nonumber\\
\left.\frac{\langle \Delta J_b\rangle}{\langle J_b\rangle}\right\vert_{tot}
&=& -\frac{U_0}{E} \left(1 - \frac{I_{4b}}{I_2}\right)\, ,
\end{eqnarray}
where we have to keep in mind that this is the variation of the action during one turn
in the storage ring which has the duration $T_0.$ Thus we find for the damping times
$\tau_a$ and $\tau_b$ for the two normal modes
\begin{eqnarray}\label{eq:taus}
\frac{d\langle J_a\rangle}{dt}&\approx&\frac{\langle \Delta J_a \rangle}{T_0}
= -\frac{U_0}{T_0E} \left(1 - \frac{I_{4a}}{I_2}\right)\langle J_a\rangle
= -\frac{2}{\tau_a}\langle J_a\rangle\nonumber\\
\frac{d\langle J_b\rangle}{dt}&\approx&\frac{\langle \Delta J_b \rangle}{T_0}
= -\frac{U_0}{T_0E} \left(1 - \frac{I_{4b}}{I_2}\right)\langle J_b\rangle
= -\frac{2}{\tau_b}\langle J_b\rangle
\end{eqnarray}
and we recover the well-known expression~\cite{SANDS} for the damping time, but generalized to
describe the damping of the normal modes. Note that the conventional damping time refers to the
damping of the amplitude of betatron oscillations, whereas in eq.~\ref{eq:taus} we calculate the
reduction of the action variable, which accounts for the factor two in the definition of the
damping time.
The terms in brackets are commonly called the damping partition numbers $\cal J$, which tell how the
total damping is distributed among the two transverse and the longitudinal mode.
\beq
{\cal J}_{a,b} = 1-\frac{I_{4a,b}}{I_2}.
\eeq
As proven under very general conditions, which include coupling~\cite{ROBINSON}, the total
damping is constant, and as a consequence the damping partition numbers always sum up to~4. For the longitudinal damping partition and the three damping times we thus get:
\beq\label{eq:dparte}
{\cal J}_e = 4 - {\cal J}_a-{\cal J}_b\qquad\mathrm{and}\qquad \tau_{a,b,e}=\frac{2ET_0}{U_0{\cal J}_{a,b,e}}
\eeq
Moreover our finding $I_{4a}+I_{4b}=I_4$ implies that ${\cal J}_e$ does not change with transverse
coupling.\\

\section{Emittance}
In sections \ref{sec:ex} and \ref{sec:damp} we derived the generalized quantum excitation and damping integrals $I_5$ and $I_4$ including coupling. The other three integrals $I_1$, $I_2$, $I_3$ do not depend on coupling.
The integral $I_1=\oint h(s) D_x (s) \, ds$ determines the energy dependent path length (momentum compaction). Here only the dispersion in the plane of deflection is required; the dispersion in the orthogonal plane does not affect the path length in first order. A coordinate rotation may perform the transformation to entry or exit of a bend oriented differently.
The integrals $I_2=\oint h(s)^2\,ds$ and $I_3=\oint |h(s)|^3\,ds$ do not depend on any optical functions and are therefore not affected by coupling.\\

The equilibrium normal-mode emittances, which we identify as the ensemble
average of the action variables $\eps_{a,b}=\langle J_{a,b}\rangle$, are given by the
balance of heating, as described by eq.~\ref{eq:heating} and damping in eq.~\ref{eq:taus}
with the result
\begin{equation}\label{eq:emitt}
\eps_a = C_q \gamma^2 \frac{I_{5a}}{I_2-I_{4a}}
\quad\mathrm{and}\quad
\eps_b = C_q \gamma^2 \frac{I_{5b}}{I_2-I_{4b}}\ .
\end{equation}
These are the same expressions found for uncoupled rings with the exception that
here we use the generalized radiation integrals that describe the projections of
excitation and damping on to the normal modes.
\par
Turning back to the general coupled case and given the normal-mode emittances and
the decomposition of the transfer matrix it is straightforward to determine the beam
matrix in real space from that in normalized phase space. With the normal-mode
emittances $\eps_a$ and $\eps_b$ the beam sigma matrix in normalized phase space
$\tilde\sigma$ is given by
\begin{equation}
\tilde \sigma =\left(\begin{array}{cccc}
\eps_a & 0 & 0 & 0 \\
0 &\eps_a & 0 & 0 \\
0 & 0 & \eps_b & 0 \\
0 & 0 & 0 &\eps_b
\end{array}\right)
\end{equation}
and the transfer matrix $S=T^{-1}{\cal A}^{-1}$ transports the particles from normalized
phase space back to real space such that the sigma matrix $\sigma$ becomes
\begin{equation}\label{eq:sigmamat}
\sigma=S\tilde\sigma S^T
\end{equation}
and the projected emittances can be extracted from $\sigma$ by taking the
determinant of the $2\times 2$ blocks on the diagonal which are the square
of the projected emittances~$\hat\eps_{x/y}.$
\par
A parameter quantifying the coupling between the planes can be introduced either by
taking the ratio of the the normal-mode emittances $\tilde\kappa=\eps_b/\eps_a$ or
the projected emittances $\kappa=\hat\eps_y/\hat\eps_x.$ Which to take in a particular
case is a matter of taste or convenience.\\

Since $I_3$ is not affected by coupling, the common formula for the energy spread is reproduced  using the  longitudinal damping partition from eq.~\ref{eq:dparte}:
\beq\label{eq:espread}
\sigma_e^2 =C_q\gamma^2\frac{I_3}{I_2 {\cal J}_e} = C_q\gamma^2 \frac{I_3}{2I_2+I_{4a}+I_{4b}}
= C_q\gamma^2 \frac{I_3}{2I_2+I_{4}}
\eeq
Finally the r.m.s. bunch length follows with $I_1$ from the energy spread and the synchrotron tune $Q_s$ (for $Q_s << 1$):
\beq\label{eq:blength}
\sigma_s = \frac{I_1}{2\pi Q_s} \sigma_e.
\eeq
Eqs.~\ref{eq:sigmamat}--\ref{eq:blength} give the 6-dimensional equilibrium beam parameters at any location in a coupled lattice. They are determined by the generalized radiation integrals together with transverse (magnet gradients) and longitudinal (RF voltage) focusing.

\section{\label{sec:orbit}Orbit distortion}
Magnet misalignments, active steerers or a beam energy offset cause a deviation of the beam centroid (orbit) from the reference path, which is defined by the multipole centers. As a consequence the beam experiences transverse fields and gradients in all multipoles, i.e. all magnets act as small bending magnets and thus contribute to the radiation integrals via feed-down of fields and gradients at the orbit position.
We do not explicitely consider skew multipoles instead they are realized by rotations of regular multipoles.
\par
Defining $X(s)$, $Y(s)$ as the local orbit relative to the reference trajectory, the coordinates of a particle oscillating around the orbit are $(X+x)$ and $(Y+y)$. Keeping only linear, regular multipoles and using again curvatures (inverse bending radii) we define local curvatures
\beq\label{eq:curvatures}
h_X =   \left. \frac{B_y}{(B\rho)} \right|_{x,y=0} = h_0 + k X \qquad
h_Y = - \left. \frac{B_x}{(B\rho)} \right|_{x,y=0} = - k Y,
\eeq
with $h_0$ the curvature of the reference trajectory and $k$ the gradient.

Generalizing the preceding treatments to an arbitrary orbit, we first realize that the radiation integrals $I_2$, $I_3$, $I_5$ contain only the quantum emission term which is given by the absolute curvature. At the orbit it is given by
\beq\label{eq:habsorb}
|h|=\sqrt{h_X^2+h_Y^2} = \sqrt{ h_0^2+2h_0kX+k^2(X^2+Y^2)}.
\eeq

For the damping integral $I_4$ the variation of path length with offsets $x,y$ to the displaced orbit at $X,Y$ has to be considered. Using local orbit curvatures $h_X$, $h_Y$ the path length is approximated by
\beq\label{eq:pathho}
dl
\approx ds (1+h_X x + h_Y y),
\eeq
if we keep only linear terms in coordinates. We use $dl$ as a generic symbol for a path length element and not as global variable.

As derived in sec.~\ref{sec:damp}, $I_4$ depends on the radiated power as a function of transverse position, which is proportional to the local field squared, see eq.~\ref{eq:Pgamma}: the radiation on the orbit depends on the field at the orbit and the energy of the orbit.
So the average relative energy of the emitted photons from eq.~\ref{eq:ebarsector0} is given by
\beq
\Delta \bar{u} = \frac{P_{\gamma}}{E} \frac{dl}{c} =
\hat{C} \left[ (h_X+kx)^2 +(h_Y-ky)^2\right]\cdot
(1+h_X x + h_Y y)\, ds
\eeq
with $\hat{C}=C_{\gamma}E^3/(2\pi)$ and the path length element from eq.~\ref{eq:pathho}, now also including the relative offset to the orbit. The first bracket corresponds to the squared magnetic field at the location of the particle.
Keeping only linear terms in the coordinates $x,y$ we get
\begin{eqnarray}
\Delta \bar{u} & = & \hat{C} \left( C_0+C_x x +C_y y\right)\,ds ,\quad\mbox{with} \label{eq:duorb}\\
C_0 & = & 
|h|^2 \label{eq:orbco}\\
C_x & = & 
h_X\,( |h|^2 +  2k) \label{eq:orbcx} \nonumber \\
C_y & = &  
h_Y\,(|h|^2-2k) \nonumber
\end{eqnarray}
where we used eq.~\ref{eq:curvatures} and eq.~\ref{eq:habsorb}.
We insert \ref{eq:duorb} into eq.~\ref{eq:jabdefinition} describing the change of betatron amplitudes due to radiation emission. Further we insert eq.~\ref{eq:sum1i} expressing $x$ in normalized coordinates $\tilde{x}_i$, and the corresponding expression for $y$:
\beq
\Delta J_a = \left( \tilde{x}_1\tilde{D}_1+\tilde{x}_2\tilde{D}_2\right)
\hat{C} \left[ C_0+C_x \sum_{i=1}^4 S_{1i}\tilde x_i +C_y \sum_{i=1}^4 S_{3i}\tilde x_i \right] ds\, .
\eeq
Averaging over betatron phases (cf. treatments following eq.~\ref{eq:jasector}) and performing the corresponding procedure for $J_b$ results in
\begin{eqnarray}
\langle \Delta J_a\rangle  &=& \hat{C} \left[
C_x \left( S_{11} \tilde D_1 + S_{12} \tilde D_2\right) +
C_y \left( S_{31} \tilde D_1 + S_{32} \tilde D_2\right)
\right] ds\, \langle  J_a\rangle \nonumber\\
\langle \Delta J_b\rangle  &=& \hat{C} \left[
C_x \left( S_{13} \tilde D_3 + S_{14} \tilde D_4\right) +
C_y \left( S_{33} \tilde D_3 + S_{34} \tilde D_4\right)
\right] ds\, \langle J_b\rangle
\end{eqnarray}
With the relations between normalized, decoupled and real space dispersions (see eqs.~\ref{eq:s12d}, \ref{eq:s34d} and the corresponding ones for $D_y$, ${\cal D}_b$) we get
\begin{eqnarray}
\frac{\langle \Delta J_a\rangle }{\langle J_a\rangle } &=& \hat{C} \left[
C_x g {\cal D}_a + C_y \left( D_y-g{\cal D}_b\right)\right] ds \nonumber\\
\frac{\langle \Delta J_b\rangle }{\langle J_b\rangle } &=& \hat{C}\left[
C_y g {\cal D}_b + C_x \left( D_x-g{\cal D}_a\right)\right] ds
\end{eqnarray}
On the reference orbit ($X=Y=0$) we get $C_x=h_0 (2k+h_0^2)$, $C_y=0$ and recover eq.~\ref{eq:jae4}. For a pure quadrupole we simply set $h_0=0$.\\

If the magnet has a rotated edge of angle $\theta$, then the path is shortened by
$\Delta L =  -x \tan\theta$, see Fig.~\ref{fig:bendgeo}~--  now the orbit at $X$ is the reference.
Using $\Delta L$ instead of $dl=ds(1+h_Xx+h_Yy)$ we repeat the calculations
and arrive at an expressions that corresponds to eq.~\ref{eq:orbco} for the edge angles:
\begin{equation}
\bar{C}_0  =  0  \qquad
\bar{C}_x  =  -  \tan\theta\, |h|^2 \qquad
\bar{C}_y  =  0 \label{eq:orbcedge}
\end{equation}
The expressions are the same like on-axis, cf. \ref{eq:dampaedge}.
\par
The radiation integrals also include $I_1$, which actually is not related to radiation but describes the energy dependent path length.
Including an energy offset $u$ the path length element from eq.~\ref{eq:pathho} becomes
\[ dl=ds(1+(h_X D_x +h_Y D_y) u )\]
Any higher orders of path length (i.e. non-linear momentum compaction) are contained in the local dispersion $D_x$.
\par
Collecting our results we obtain Table~\ref{tab:radi} giving a summary of the contributions to the radiation integrals from a general (combined function) bending magnet with (constant) curvature $h_0$, quadrupole strength $k$ and edge angles $\theta_{1,2}$ including coupling and orbit distortions. Our results agree with prior work by Sagan as documented in \cite{BMADMAN}.\\

\begin{table}[t]
\caption{\label{tab:radi}Radiation integrals for a general bending magnet with curvature $h_0$ and focusing strength $k$ including coupling and orbit distortions.}
\begin{eqnarray*}
I_1 & = & \int (h_X D_x +h_Y D_y)\, ds \qquad
I_2 = \int |h(s)|^2\,ds \qquad
I_3 = \int |h(s)|^3\,ds \\
I_{4a} &=& \int
\left[ C_x(s) g {\cal D}_a(s) + C_y(s) \left( D_y(s)-g {\cal D}_b(s)\right) \right]\,ds
-\sum_{i=1,2} \tan\theta_i \bar{C}_{xi} g {\cal D}_{ai} \\
I_{4b} &=& \int
\left[ C_y(s) g {\cal D}_b(s) + C_x(s) \left( D_x(s)-g {\cal D}_a(s)\right) \right]\,ds
-\sum_{i=1,2} \tan\theta_i \bar{C}_{xi}
\left( D_{xi}-g {\cal D}_{ai} \right)  \\
I_{5a} &=& \int \tilvec{D}_a^T (s) \cdot \tilvec{D}_a (s)\,  |h(s)|^3\,ds \qquad
I_{5b} = \int \tilvec{D}_b^T (s) \cdot \tilvec{D}_b (s)\, |h(s)|^3\,ds
\end{eqnarray*}
\[
{\rm with} \qquad h_X(s) = h_0+k X(s) \qquad h_Y(s) = -k Y(s) \qquad |h(s)| = \sqrt{(h_X(s))^2+(h_Y(s))^2}
\]
\[
C_x (s)  =  h_X(s)\,(|h(s)|^2 + 2k(s) ) \qquad
C_y (s)  =  h_Y(s)\,(|h(s)|^2 - 2k(s) ) \qquad
\bar{C}_x  =  |h_i|^2
\]
\end{table}

In the calculations we made the following assumptions and simplifications:

(a) The integrals extend along the displaced orbit. In a magnet of cylindrical symmetry with design curvature $h_0$, the integration path element thus becomes
\[
dl = ds\,(1+h_0 X(s)).
\]
If the magnet further has a rotated edge of angle $\theta$, then the path is shortened by
\[
\Delta L = -X \tan\theta, 
\]
see Fig.~\ref{fig:bendgeo}. An integral over a function $f(s)$ thus becomes
\[
\int_{\rm path} f(s)\, dl = 
\int_{\Delta L_1}^{L-\Delta L_2} f(s)\,(1+h_0X) \,ds
\approx \int_0^L f(s)\,(1+h_0X) \,ds - \sum_{i=1,2} f(s_i) \tan\theta_i\, X_i.
\]
Basically this path length correction has to be applied to all radiation integrals.
However, for storage rings operating at several GeV beam energy, the radius of curvature $\rho=1/h_0$ is on the orders of meters, while the orbit excursion $X$ is typically on the order of some millimeter. As a consequence, the path length modifications are a per-mill effect and may be neglected. For rectangular bends, as commonly used, where $\theta_1=\theta_2=h_0L/2\ll 1$ the two effects, lengthening of the path in the arc and shortening due to cut off by rotated edges, cancel anyway.

(b) It would be straightforward to extend the treatment to non-linear multipoles too, however the contribution to the integrals is rather small:
the field increases quadratically (sextupole) or at even higher power with the distance from the axis, and the radiation integrals scale with the second or third power of the field. Even if the sextupoles are rather strong, like in a low-emittance storage ring, their contribution to the integrals is on a per-mill level compared to the quadrupole contribution~-- which in itself is a fraction of the bending magnet contribution~-- and thus may well be neglected.

(c) We only consider the design edge rotation (around the $y$-axis) of a bending magnet and neglect additional edge effects due to orbit slopes $X'$, $Y'$, because these angles are usually small compared to bending and edge angles. Thus the edge related path shortening does not depend on the vertical coordinates. Note that we only considered horizontal bending magnets, which may be sandwiched between rotations to model vertical or slanted deflections.

(d) During propagation the effect known as ``mode flip'' may occur: the normal mode decomposition locally may result in one or two solutions for the matrix $T$ of Eq.~\ref{eq:tmatrix}. In strongly coupled lattices it can happen, that the solution in use does not exist anymore at the next element, and the other solution has to be used to propagate further, as described in \cite{SAGAN}. In this case we proceed by adding the $a$-integrals to the $b$-sum integral of the ring and vice versa, until the originally used solution exists again. However, since we restricted ourselves to regular magnets all changes of coupling (as expressed by the parameter $g$) occur at rotations, which are inserted before and after the magnets if needed, and no mode flip can occur during integration inside the magnet.

\section{Applications}

Several simulation codes are in use, e.g. MAD-X \cite{MADX}, elegant \cite{ELEGANT}, AT \cite{AT}, BMAD \cite{BMAD}, Tracy \cite{TRACY}, which incorporate different coupling formalisms \cite{SLIM}--\cite{SAGAN} and calculate the equilibrium parameters for a given storage ring lattice.
The code OPA \cite{OPA} emerged from the former code OPTIK, which was written by K. Wille and colleagues at Dortmund university in the 1980's. The philosophy of the code was different providing an interactive, visual tool to build up a lattice from scratch like playing with LEGO\textsuperscript{TM} blocks, while restricting the models to elementary magnet types and functionalities.

For interactive lattice design from scratch it is convenient, although physically meaningless, to calculate radiation integrals even if it is not yet known if and how the lattice will be terminated or closed, i.e. irrespective of the existence of a periodic solution. For this purpose a forward propagation using analytical expressions for the integrals is required, which starts from initial beam parameters defining orbit, normal mode betas, coupling and dispersion. If a periodic solution exists, it will deliver the initial parameters and the same calculation can be applied.
The OPA-code is based on this concept. Extending it to coupled lattices partially motivated the work described here.

The presented method to calculate the radiation integrals for the normal modes permits a
straightforward implementation of the formulae in Table \ref{tab:radi}:
We assume that curvature $h$ and gradient $k$ are piecewise constant, then the transfer matrix $R$ and the vector $\vec{V}$ of dispersion production in eq.~\ref{eq:rmatrix} contain trigonometric or hyperbolic functions (to be found in any text book on accelerators, e.g.\cite{SYPHERS}, \cite{HANDBOOK}, \cite{SYLEE}), whereas otherwise analytic solutions can only be found for a few special cases of longitudinal field variation.

Constant $h$, $k$ parameters can be extracted from the radiation integrals, and we are left with sums of definite integrals over multiple products of trigonometric or hyperbolic functions, which have elementary solutions. These solutions become rather lengthy and are not instructive at all,  therefore we do not show them here. But they can be evaluated and exported as program code by a symbolic code, which was the first and most efficient implementation.
Solving the integrals numerically by dividing the magnets into sufficiently small slices instead of evaluating the analytical solutions is less elegant and computationally more expensive, but easy and straightforward to implement and to extend further, in order to also include orbit distortions, longitudinal field variations etc.

Linear beam optics is calculated from the local transfer-matrix, which is the  Jacobian of the non-linear map at the orbit position. The orbit itself is the fixpoint of the one-turn map in a periodic system, or just the propagation of initial conditions including all multipoles in a single-pass system. Off-axis down-feeds have to be included in linear optics, like the additional dispersion production on the displaced orbit, $\Delta\vec{D}(s) = d\vec{X}(s)/du |_{u=0}$, and local quadrupole and skew quadrupole gradients from non-linear multipoles.

Implementation details and practical issues will be described elsewhere~\cite{OPA}.
In the following we will show four example applications:\\

\begin{figure}
\centering
\includegraphics[width=72mm]{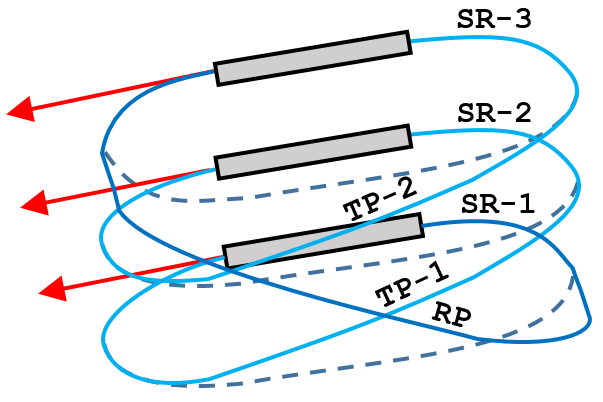}
\caption{\label{fig:spiral1}Schematic layout of the Spiral-COSAMI lattice (from \cite{SPIRAL}).}
\end{figure}

\begin{figure}
\centering
\includegraphics[width=\textwidth]{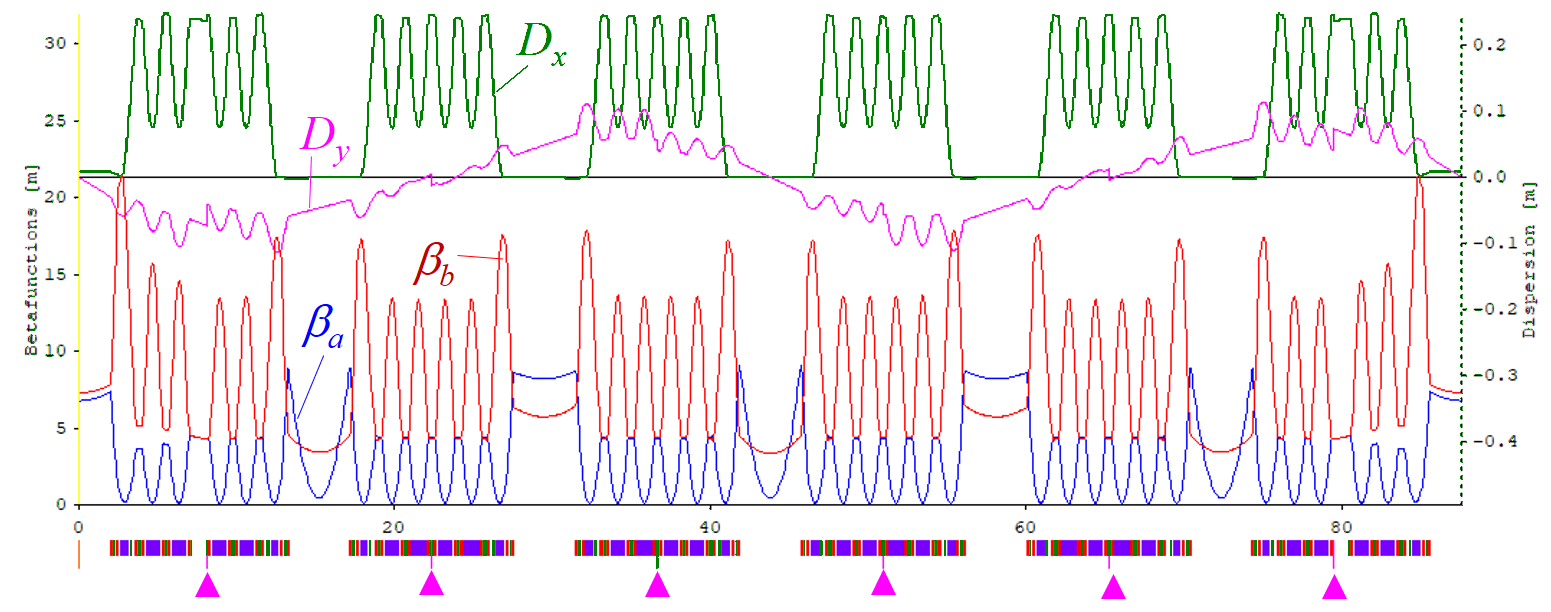}
\caption{\label{fig:spiral2}Optics of the Spiral COSAMI lattice showing normal mode beta functions (left axis) and physical dispersions (right axis). The triangles at the bottom indicate points of rotations by angles $\{ -\beta, \alpha, \alpha, \alpha, \alpha, -\beta\} $ with $\alpha = 3.86^{\circ}$, $\beta = 7.69^{\circ}$}
\end{figure}

Example~1: The Spiral-COSAMI storage ring is a machine for industrial application of extreme ultraviolet light \cite{SPIRAL}. It is build from three vertically stacked turns as sketched in Fig.~\ref{fig:spiral1}. The tilting of the arcs introduces coupling and excites a wave of vertical dispersion as shown in Fig.~\ref{fig:spiral2}. The figure shows the normal mode beta functions which almost coincide with the projections to physical $x,y$ space since coupling is low. The physical dispersions also shown in the figure are defined in the local coordinate system and thus show discontinuities at locations of coordinate rotation. The normal mode equilibrium emittances of this lattice amount to 3.43/0.23~nm at 430~MeV. This result was confirmed by the code TRACY~\cite{TRACY}, which employs a different method of emittance calculation based on the 6-dimensional periodic sigma-matrix~\cite{CHAO}.\\

\begin{figure}
\centering\noindent
\includegraphics[width=\textwidth]{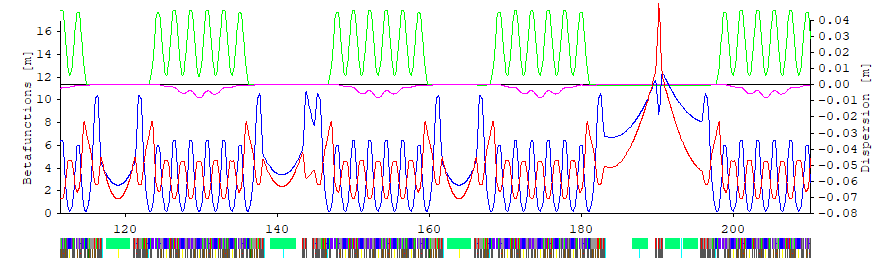}
\caption{\label{fig:sls1} Normal normal mode beta functions (blue $a$-mode, red $b$-mode) and dispersions (green horizontal, magenta vertical) for setors 6 to 9 of the SLS~2.0 lattice loaded with insertion devices.}
\end{figure}

\begin{figure}
\centering\noindent
\includegraphics[width=8cm]{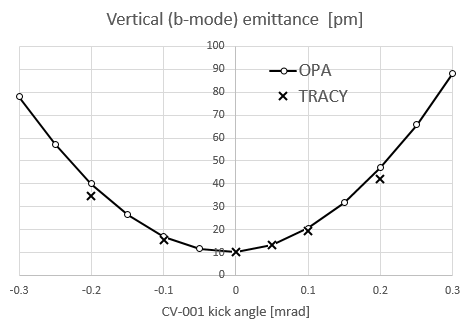}
\caption{\label{fig:sls2} Comparison of OPA- and TRACY calculations of vertical emittance due to orbit distortion, if one vertical corrector is excited (located at $\beta_y=7.6~m$).}
\end{figure}

Example~2: The lattice for the upgrade of the Swiss Light Source, SLS~2.0~\cite{TDR} has a natural horizontal emittance of 158~pm at 2.7~GeV. 264 skew quadrupoles are set to generate closed bumps of vertical dispersion in the arcs in order to create 10~pm of $b$-mode emittance, which is a compromise between beam life time due to Touschek scattering and photon beam brightness.
Due to the rather low coupling the vertical emittance basically is given by the $b$-mode emittance.
Closing all undulators to minimum gaps increases the radiated power as given by the $I_2$-integral while little affecting the other integrals, and thus reduces the normal mode emittances to 134~pm and 7.7~pm. Scaling all skew quadrupoles by the same factor proportional to $\sqrt{I_2}$ keeps the $b$-mode emittance at 10~pm for the fully loaded lattice.
Fig.~\ref{fig:sls1} shows the normal mode beta functions $\beta_a$, $\beta_b$ and the physical dispersions $D_x$, $D_y$ for one super-period of the fully loaded, ideal lattice.
Fig.~\ref{fig:sls2} demonstrates for the unloaded lattice how the vertical emittance increases due to orbit distortion when exciting one vertical corrector. We tracked the residual discrepancy
between the results from TRACY and OPA to slightly different methods to calculate the
off-axis dispersion.\\

Example~3: As an example of an extremely coupled machine we designed a ring in which we can
continuously adjust the coupling between an uncoupled configuration and a
M\"{o}bius configuration~\cite{TALMAN}, where the transverse planes are exchanged after
one turn. This racetrack-shaped ring~\cite{VZM} is based on 10\,m long FODO cells
with $90\degree$ phase advance in both planes. In the two arcs the phase advance in
the vertical plane is slightly reduced in order to split the integer tunes, while
the phase advance in the horizontal plane remains at $90\degree$, which allows us to implement
a dispersion suppressor with four half-length dipoles. The two straight sections
consist of six $90\degree$ FODO cells. One is used to adjust tunes and the other
implements a M\"{o}bius tuner with three skew quadrupoles placed in the middle
of corresponding drifts spaces in consecutive cells~\cite{MOBIUS}. This
ensures that the phase advance
between consecutive skew quadrupoles is $90\degree$ in both planes. If we now choose
the focal length of the skew quadrupole to be equal to the beta function in the
middle of the straight, the transfer matrix for this section has $2\times 2$
blocks with zeros on the diagonal and non-zero entries in the off-diagonal blocks.
In other words, it exchanges the transverse planes. Adjusting the excitations
of the three skew quadrupoles by the same factor allows us to continuously vary
the coupling between an uncoupled ring and one in the M\"{o}bius configuration.
\par
\begin{figure}[tb]
\begin{center}
\includegraphics[width=0.9\textwidth]{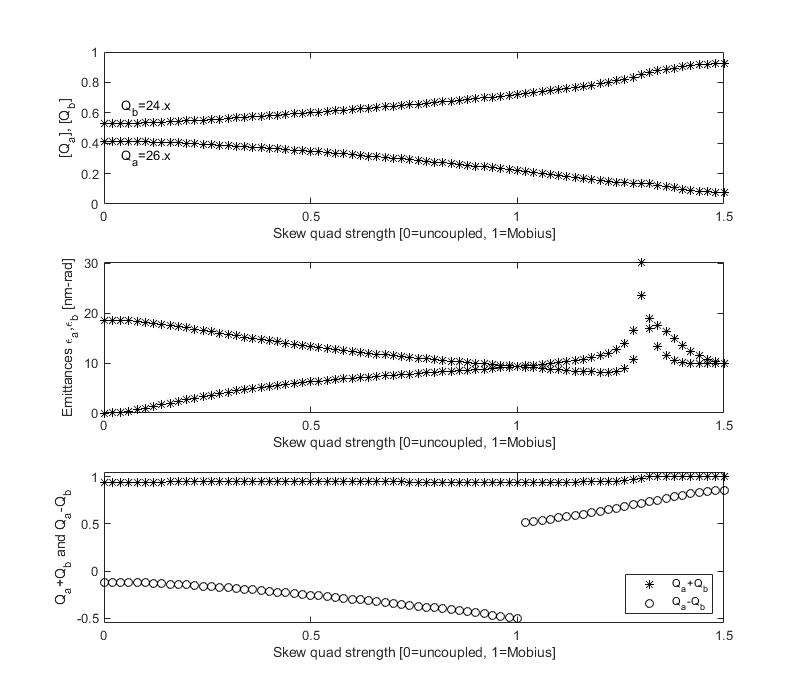}
\end{center}
\caption{\label{fig:mobius3}The fractional tunes (top) and the emittances (middle)
  of the M\"{o}bius ring (top) as a function of the excitation of the skew quadrupoles. The
  bottom panel shows the sum and difference of the tunes.}
\end{figure}
The upper panel in Figure~\ref{fig:mobius3} shows the fractional tunes for the ring
as the excitation of the skew quadrupoles is increased from zero to the M\"{o}bius
configuration and up to 1.5 times that excitation. The integer
part of the tune is indicated on the left-hand side of the plot. We see that increasing the excitation
of the skew quadrupoles ``pushes the fractional tunes apart'', which makes it
beneficial to place one tune above the half-integer and the other below.  This
prevents the crossing of the half integer with the ensuing instability for one of
the tunes. The middle panel shows the emittances
calculated with Eq.~\ref{eq:emitt}. We observe that increasing the excitation of the
skew quadrupoles to the M\"{o}bius configuration reduces one of the emittances and
increases the other one such that they become equal. The lower panel shows that
the difference between the fractional part of the tunes becomes half-integer at
this point. Increasing their excitation further causes both emittances to increase
dramatically. The plot on the lower panel provides an explanation; the sum of the
tunes $Q_a+Q_b$ becomes an integer and the system crosses a sum resonance shown by
the asterisks in the lower panel from Figure~\ref{fig:mobius3}. And on a sum
resonance, the emittances can become arbitrarily large, because only their
difference is bounded~(Sec.2.1.3 in \cite{HANDBOOK}).\\ 

\begin{figure}
\centering
\includegraphics[width=\textwidth]{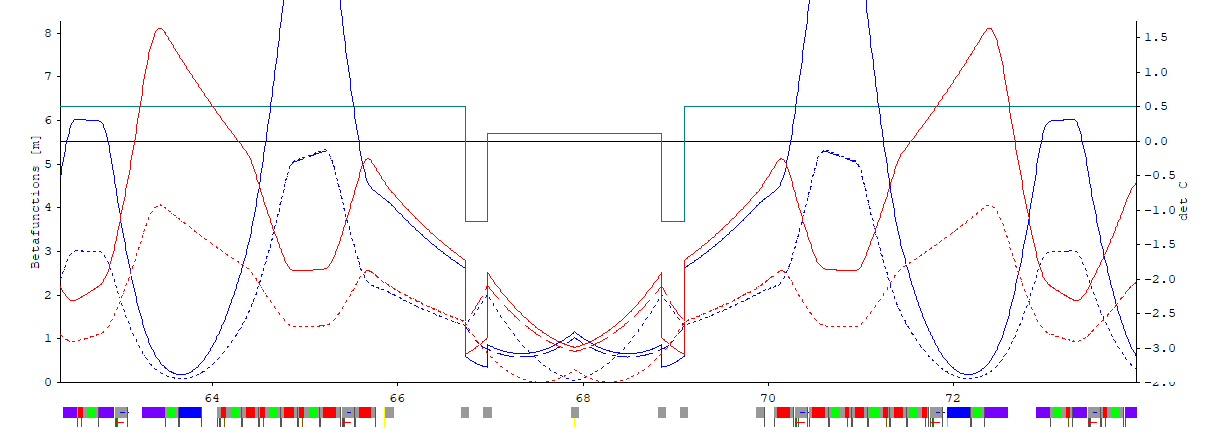}
\caption{\label{fig:moin}
M\"{o}bius insertion in straight 4 of the SLS 2.0 storage ring. Solid lines show the normal mode betafunctions (blue $a$-mode, red $b$-mode), dashed and dotted lines the projected beta functions $\beta_{xa}, \beta_{yb}$ and $\beta_{xb}, \beta_{ya}$ (blue $\beta_x$, red $\beta_y$). The square line shows the determinant of the coupling matrix $\det C =1-g^2$ (axis at right).}
\end{figure}

Example~4: A M\"{o}bius insertion in a light source is an option to lower the bunch density in order to minimize emittance growth and particle losses (Touschek effect) due to intra-beam scattering. Furthermore, some beam lines prefer photon beams of almost round cross section obtained in this way.
Our last example demonstrates application to the SLS~2.0 lattice:
The M\"{o}bius insertion could be realized in one of the of the short straight sections.
It is made from five skew and two regular quadrupoles~\cite{MOBIUS}. Figure~\ref{fig:moin} shows the optical functions in magnification: normal mode beta functions jump at the location of the (thin) skew quadrupoles, whereas their projections to $x$- and $y$-axes, which correspond to the physical beam envelopes, vary continuously. Outside the M\"{o}bius insertion the projections from the $a$- and $b$-modes to $x$, resp. to $y$ coincide, indicating full coupling at $g=\sqrt{1/2}$. As a consequence the radiation integrals for both modes are identical, and so are the damping times and the emittances. The uncoupled lattice without insertion devices has a natural emittance of $\epsilon_{xo}=158$~pm and damping partitions of ${\cal J}_{xo}=1.828$, ${\cal J}_{yo}=1.000$. The M\"{o}bius lattice has ${\cal J}_a={\cal J}_b=({\cal J}_{xo}+{\cal J}_{yo})/2 = 1.414$ and  emittances of $\epsilon_a=\epsilon_b=\epsilon_{xo}/(1+{\cal J}_{yo}/{\cal J}_{xo})=102$~pm. These result in projected emittances of same amount, $\epsilon_x=\epsilon_y= 102$~pm outside the insertion, whereas projected emittances are larger inside the insertion.

\section{Conclusion}
We found a generalization of the synchrotron radiation integrals that are
responsible for the emittance growth in coupled lattices. It turned out that the
resulting methodology is rather intuitive. It consist of projecting the four dimensional
coupled dispersion from eq.~\ref{eq:disp} onto the normal modes that are constructed
from the Edwards-Teng formalism. Then the emittance growth integral is given through the
normal-mode beta functions and dispersion~$\tilvec{D}$. The same method also worked to
project the effect of damping due to the emission of synchrotron radiation on to the
normal modes. The formalism allows off-axis contribution to be included, and implementation in a beam dynamics code is straightforward. In this way the formalism allows to introduce the coupling of the normal-mode emittances in a natural way.

\section*{Acknowledgements}
We would like to thank Masamitsu Aiba (PSI) for setting up the M\"{o}bius insertion in the latest SLS~2.0 lattice, and for useful discussions, and Michael B\"{o}ge (PSI) and Bernard Riemann (PSI) to perform cross-checks of our example applications using the TRACY code. We further gratefully acknowledge helpful comments from David C. Sagan (Cornell University).

\bibliography{vzascoup}
\bibliographystyle{apsrev}

\end{document}